\documentclass[prl,twocolumn,preprintnumbers,tightenlines,superscriptaddress]{revtex4}
\usepackage{graphicx}
\usepackage{amssymb}
\usepackage{amsmath,mathrsfs,verbatim}
\usepackage{times}
\usepackage{latexsym}
\usepackage{color}
\usepackage{graphicx}
\usepackage{cancel}
\newcounter{sec}
\def\beq{\begin{equation}}
\def\eeq{\end{equation}}
\def\bea{\begin{eqnarray}}
\def\eea{\end{eqnarray}}
\def\bit{\begin{itemize}}
\def\eit{\end{itemize}}
\def\ben{\begin{enumerate}}
\def\een{\end{enumerate}}
\def\Nef{N_{\rm eff}}
\begin{document}

\title{Visualizing Invisible Dark Matter Annihilation with the CMB and Matter Power Spectrum}

\author{Yanou Cui}
\email{yanou.cui@ucr.edu}
\affiliation{Department of Physics and Astronomy, University of California, Riverside, California 92521, USA}
\author{Ran Huo}
\email{huora@ucr.edu}
\affiliation{Department of Physics and Astronomy, University of California, Riverside, California 92521, USA}

\date{\today}

\begin{abstract}
We study the cosmological signatures of Invisibly Annihilating Dark Matter (IAnDM), where DM annihilates into dark radiation particles that are decoupled from the Standard Model (SM). In the simple benchmark model we consider here, such invisible annihilation determines the relic abundance of DM via dark thermal freeze-out.  We demonstrate that IAnDM may reveal itself through observable, novel signatures that are correlated:  scale-dependent $\Delta N_\text{eff}$ (number of extra effective neutrinos) in the Cosmic Microwave Background (CMB) spectrum due to DM residual annihilation, while the phase of acoustic peaks shifts towards the opposite direction relative to that due to SM neutrinos, resembling the effect due to scattering (fluidlike) thermal dark radiation; in addition, IAnDM induces modifications to the matter power spectrum that resemble those due to warm dark matter. Current data are sensitive to IAnDM with masses up to $\sim200$~keV, while future observations will improve this reach, especially if the late-time DM annihilation cross section is enhanced relative to the standard thermal value, which can be realized in a variety of scenarios. This work also demonstrates a new possibility of realizing thermal sub-MeV DM with observable signals.
\end{abstract}

\maketitle

\stepcounter{sec}
{\bf \Roman{sec}. Introduction\;}

Over the past two decades, we have seen overwhelming gravitational evidence for the existence of dark matter (DM), which constitutes $80\%$ of the total matter density in our Universe today. Nevertheless, the non-gravitational, particle nature of DM remains mysterious. Among the many theoretical candidates for DM, the Weakly Interacting Massive Particle (WIMP) is a well-motivated scenario that has played a central role in guiding the experimental searches for DM. The key process in WIMP models is the annihilation of DM into Standard Model (SM) particles, which determines the relic abundance of DM after its thermal freeze-out in the early Universe. The residual DM annihilation today is potentially observable through indirect detection experiments, and via the Cosmic Microwave Background (CMB) observations which are free of astrophysical uncertainties~\cite{Padmanabhan:2005es, Madhavacheril:2013cna, Green:2018pmd}. These CMB studies focus on the effects of \textit{visible} SM states (\text{e.g.}, $e^\pm, \gamma$) on the recombination history of the Universe.

Recently, driven by the strengthening experimental constraints on WIMP DM, there has been a growing interest in dark sector scenarios, where DM resides in a hidden sector with multiple states and/or self-interactions, in analogy to the complex structure of the SM. In the simple benchmark model that we focus on here, the merit of predicting DM relic abundance from thermal freeze-out is retained, yet the DM predominantly annihilates into other dark states instead of SM particles~\cite{Pospelov:2007mp, Feng:2008mu, Cheung:2010gk, Cline:2011uu, Belanger:2011ww, Blennow:2012de, Garcia-Cely:2013wda, Agashe:2014yua, Berger:2014sqa, Chacko:2015noa, Chacko:2016kgg, Chan:2016mhw, Bringmann:2018jpr}. By decoupling the thermal relic abundance from couplings to the SM, these models are relieved from conventional DM constraints, yet they call for new strategies for detection. The most challenging case arises when DM undergoes \textit{invisible annihilation} - i.e., when DM annihilates into stable dark radiation (DR) particles that are decoupled from the SM. Can we \textit{directly} observe or constrain this apparent nightmare (yet minimal and generic) scenario of Invisibly Annihilating DM (IAnDM) with cosmological data? In this paper, we will demonstrate a positive answer to this question: while the thermal $\Delta\Nef$ of DR~\cite{Chacko:2015noa} in IAnDM can be significantly suppressed if the dark sector is much colder than the SM, an observable yet \textit{nonstandard} $\Delta\Nef$ effect can result from the nonthermal DR directly produced from DM annihilation. For IAnDM to yield detectable gravitational effects in the CMB and matter power spectrum (MPS), DM needs to be light and copious. In such a mass range, the MPS is dominantly affected by (\textit{cold}) DM free-streaming. IAnDM with masses under $\sim200$~keV is disfavored by current data, while future experiments will improve the sensitivity reach, especially if the late-time annihilation cross section is enhanced relative to the standard thermal value, which can be generally realized (see discussion in Sec.~V).

\stepcounter{sec}
{\bf \Roman{sec}. An Example Model\;}

The scenario of IAnDM can generally arise in particle physics models with thermal DM or beyond, while \textit{our main results are largely model independent}. We consider a simple example model where the SM singlet fermionic DM $\chi$ annihilates into the lighter fermionic DR particle $\psi$ through a vector current interaction ($Z'$ mediator), with an $s$-wave thermal cross section, $\langle\sigma v\rangle$, which determines the relic abundance of $\chi$, $\Omega_\chi$. The relevant Lagrangian is:
\begin{equation}
\mathcal{L}\supset i\bar{\chi}\cancel{D}\chi+i\bar{\psi}\cancel{D}\psi-m_\chi\bar{\chi}\chi-m_{\psi}\bar{\psi}\psi-\frac{1}{4}Z'_{\mu\nu}Z^{'\mu\nu}+\frac{1}{2}m_{Z'}^2Z'^2,
\end{equation}
where $D_{\mu}$ is the covariant derivative including the $Z'$ gauge interaction. The automatic self-scattering of $\chi$ or $\psi$ mediated by a moderately massive $Z'$ is generally ineffective at late times, so that $\psi$ free-streams then {(see Ref.~\cite{Chacko:2015noa} for an alternative case)}.

Without additional effective interactions, $\chi$ and $\psi$ freeze out simultaneously as $\bar{\chi}\chi\rightarrow\bar{\psi}\psi$ departs from equilibrium. In this work, we will assume that $\chi$ is the leading DM, and thus the prediction for $\Omega_{\rm DM}$ resembles that of the standard WIMP DM. This requires $\psi$ to be in the form of dark radiation, much lighter than $\chi$, and to freeze out while it is relativistic \footnote{If $\psi$ is massive enough, its relic density today can dominate over that of $\chi$ \cite{Farina:2016llk}, which is an interesting alternative scenario that we will leave for future investigation.}.  The temperature of a decoupled dark sector ($\chi, \psi$) can be much colder than the SM, depending on the reheating pattern~\cite{Berezhiani:1995am, Adshead:2016xxj} and the number of heavier states that have decoupled in each sector. The temperatures of $\chi, \psi$ also redshift differently after their freeze-out. For the interest of this work, the relevant effect can be simply parametrized by $\xi\equiv\frac{\hat{T}_f}{T_f}\leq1$ around the dark thermal freeze-out, where $\hat{T}_f$ is the dark temperature and $T_f$ is the SM one.

\begin{figure}[t!]
\includegraphics[scale=0.65]{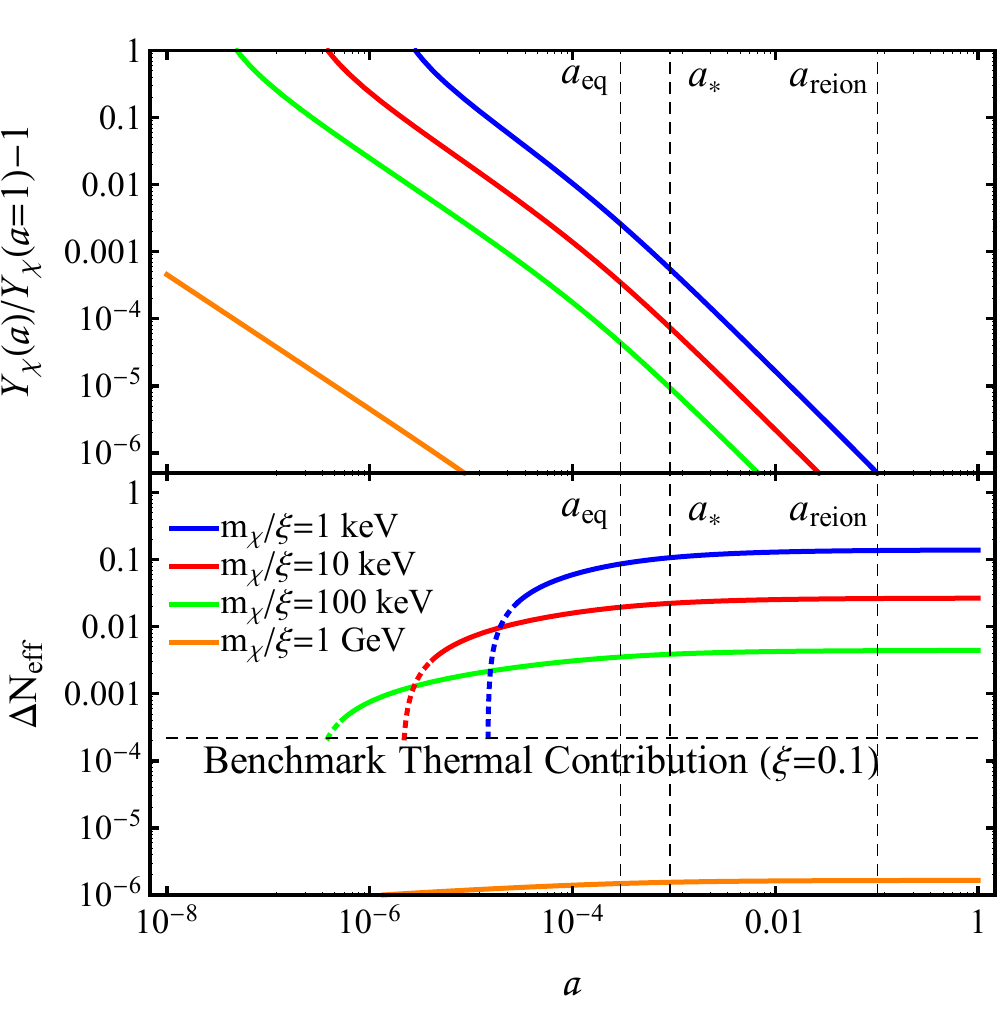}
\caption{\emph{Upper panel:} The slow depletion of $Y_\chi$ due to its residual annihilation. \emph{Lower panel:} $\Delta N_\text{eff}$ due to massless annihilation product $\psi$. $a_{\rm eq}$, $a_*$, $a_{\rm reion}$ are the expansion parameters, at the matter-radiation equality, recombination and reionization epochs, respectively. Here we include proper redshift for both RD and matter-dominated (MD) eras.}
\label{fig:FreezeOutDeltaNeff}
\end{figure}

$\Omega_\chi$ can be estimated as in Ref.~\cite{Chacko:2015noa,Feng:2008mu}. We define freeze-out temperature parameters $x_f\equiv \frac{m_{\chi}}{T_f}$, $\hat{x}_f\equiv\frac{m_{\chi}}{\hat{T}_f}=\frac{x_f}{\xi}$, and obtain
\begin{align}
x_f&\simeq \xi\ln\Big(\sqrt{\frac{45}{16}}\frac{1}{\pi^3}\frac{g_\chi}{\sqrt{g_\ast}}m_\chi M_P\langle\sigma v\rangle\xi^{\frac{5}{2}}\Big)\simeq\xi \hat{x}_f, \\
\Omega_\chi^\infty &=0.32
\Big(\frac{\hat{x}_f}{10}\Big)\Big(\frac{\sqrt{g_\ast}}{\sqrt{3.38}}\frac{43/11}{g_{\ast S}}\Big)\Big(\frac{3\times10^{-26}~\text{cm}^3~\text{s}^{-1}}{\langle\sigma v\rangle/\xi}\Big),
\label{eq:ColdDarkFreezeOut}
\end{align}
where $M_P$ is the Planck mass; $g_\chi$ is the internal degree of freedom (d.o.f.) of $\chi$; the values $g_*, g_{*s}$ are the total effective d.o.f.'s dominated by SM states; and we use the superscript $\infty$ to emphasize that the IAnDM relic abundance is the asymptotic value (after some noticeable depletion; see Fig.~\ref{fig:FreezeOutDeltaNeff}). For a light sub-MeV DM, $\hat{x}_f\sim 10$, and around the DM freeze-out $g_*=3.38,~g_{*s}=43/11$ (the SM values after neutrino decoupling). Note that a fixed $\langle\sigma v\rangle/\xi\simeq\sigma_0\equiv3\times10^{-26}~\text{cm}^3~\text{s}^{-1}$ is required to yield correct thermal DM abundance; therefore, for $\xi<1$, the required $\langle\sigma v\rangle$ is reduced by $\xi$ accordingly. In order to make sure that $\psi$ (hot relic) does not overclose the Universe, we also check the thermal relic density of $\psi$:
\begin{equation}
\Omega_{\psi\text{, th}}=\begin{cases}
\frac{m_\psi n_\psi}{\rho_c}=0.038 \Big(\frac{m_\psi}{1~{\rm eV}}\Big)g_\psi\xi^3,~~ {\rm  massive}~\psi ,\\
2.7\times10^{-4}g_\psi\xi^4,~~ {\rm  massless}~\psi ,
\end{cases}
\label{eq:Omegathm0}
\end{equation}
where the subscript ``th'' denotes the thermal relic component of $\psi$, which distinguishes itself from the nonthermal component from $\chi$ annihilation that we will focus on (denoted by the subscript ``nth''). Note that for $\xi\ll1$, $\Omega_{\psi, \text{th}}$ is strongly suppressed, even for a moderately massive $\psi$. $\xi<1$ is also required by big bang nucleosynthesis~\cite{Cyburt:2015mya} for sub-MeV DM.

\stepcounter{sec}
{\bf \Roman{sec}. Cosmological Effects of Invisible DM Annihilation: General Consideration and Analytic Studies\;} 

The nonthermal free-streaming DR $\psi$ injected from DM $\chi$ annihilation contributes an additional radiation energy component to the Universe. Assuming $m_{\psi}=0$ for simplicity, we can estimate the accumulated energy density of $\psi$ from $\chi$ annihilation, $\rho_{\psi,\rm nth}$, based on instantaneous energy conservation: $d\rho_{\psi,\rm nth}(t)={\rho_\chi}n_\chi\langle\sigma v\rangle dt$. Upon integration over time $t'$ with proper redshift factors, the accumulated $\psi$ density by the time $t$ ($a$) is
\begin{align}
\label{eq:rhoproduct}
\rho_{\psi,\rm nth}(a)=&\int_{a_i}^a\frac{\rho_{c,0}^2\Omega_\chi^2}{m_\chi a'^6}\langle\sigma v\rangle \frac{a'da'}{H_0\sqrt{\Omega_{\gamma+\nu}}}\Big(\frac{a'}{a}\Big)^4\nonumber\\
=&\Big(\frac{3H_0^2\Omega_\chi}{8\pi G}\Big)^2\frac{\langle\sigma v\rangle}{m_\chi H_0\sqrt{\Omega_{\gamma+\nu}}}\frac{\ln(\frac{a}{a_i})}{a^4}.
\end{align}
Here a deep radiation-dominated (RD) epoch is assumed, so the Hubble parameter $H(a')\equiv da'/(a'dt')=H_0\sqrt{\Omega_{\gamma+\nu}/a'^4}$ and $a_0=1$. $a_i$ represents the initial time when the \emph{net} $\psi$ production from $\chi$ annihilation becomes effective, which is around the freeze-out time, $a/a_i\sim T_f/T$. Note that in addition to the standard redshift $1/a^4$, there is a moderate \textit{log dependence} on $a$ which is different from either a fixed extra neutrino background or many other noncold DM-DR models, which is the key to the following characteristic signals.

In order to relate to the potential CMB observable, $\Delta N_\text{eff}$, we take the ratio of $\rho_{\psi,\rm nth}$ over SM neutrino density (one flavor), and find
\begin{equation}
\Delta N_\text{eff, nth}(a)=0.038\ln\Big(\frac{a}{a_i}\Big)\Big(\frac{\text{keV}}{m_\chi/\xi}\Big)\Big(\frac{\langle\sigma v\rangle/\xi}{3\times 10^{-26}~{\rm cm}^3/{\rm s}}\Big),
\label{eq:deltaNeff}
\end{equation}
where $a\approx10^{-3}$ when evaluated around the CMB epoch. Note that $\Delta N_\text{eff, nth}$ inherits the aforementioned $\ln{(a/a_i)}$ dependence. Apparently, lighter DM produces more copious $\psi$ and thus more pronounced signals. We can also see that with $\xi=1$ and thus $\langle\sigma v\rangle=\sigma_0$, an $\mathcal{O}(1)-\mathcal{O}(10)$ keV mass $\chi$ could lead to an observable $\Delta N_\text{eff}$ at current or upcoming CMB experiments~\cite{Suzuki:2015zzg, Ade:2015xua, Abazajian:2016yjj}, while for $\xi<1$ and fixed $\langle\sigma v\rangle/\xi\equiv\sigma_0$, a lighter $\chi$ is required to yield the same $\Delta N_\text{eff}$. Therefore, we will parametrize with \textit{the rescaled DM mass $m_\chi/\xi$}.

Fig.~\ref{fig:FreezeOutDeltaNeff} illustrates the physics discussed above. In the upper panel, the deviation of DM's comoving density, $Y_\chi (a)$, from its current-day value, is calculated by numerically solving the Boltzmann equation around DM freeze-out. In the lower panel, along with the novel residual annihilation induced $\Delta N_\text{eff, nth}$, the irreducible thermal contribution to $\Delta N_\text{eff}$ is also shown (relevant if $m_\psi\lesssim$ eV):
\begin{equation}
\Delta N_\text{eff, th}=3.046\frac{\rho_{\psi, \text{th}}}{\rho_{\nu, 1}}=2.2g_\psi\xi^4,
\label{eq:deltaNeffTC}
\end{equation}
where $\rho_{\nu, 1}$ is the energy density of one generation of the SM neutrino. We can see that $\Delta N_\text{eff, nth}$ can easily dominate over $\Delta N_\text{eff, th}$.

\begin{figure}[t!]
\includegraphics[scale=0.65]{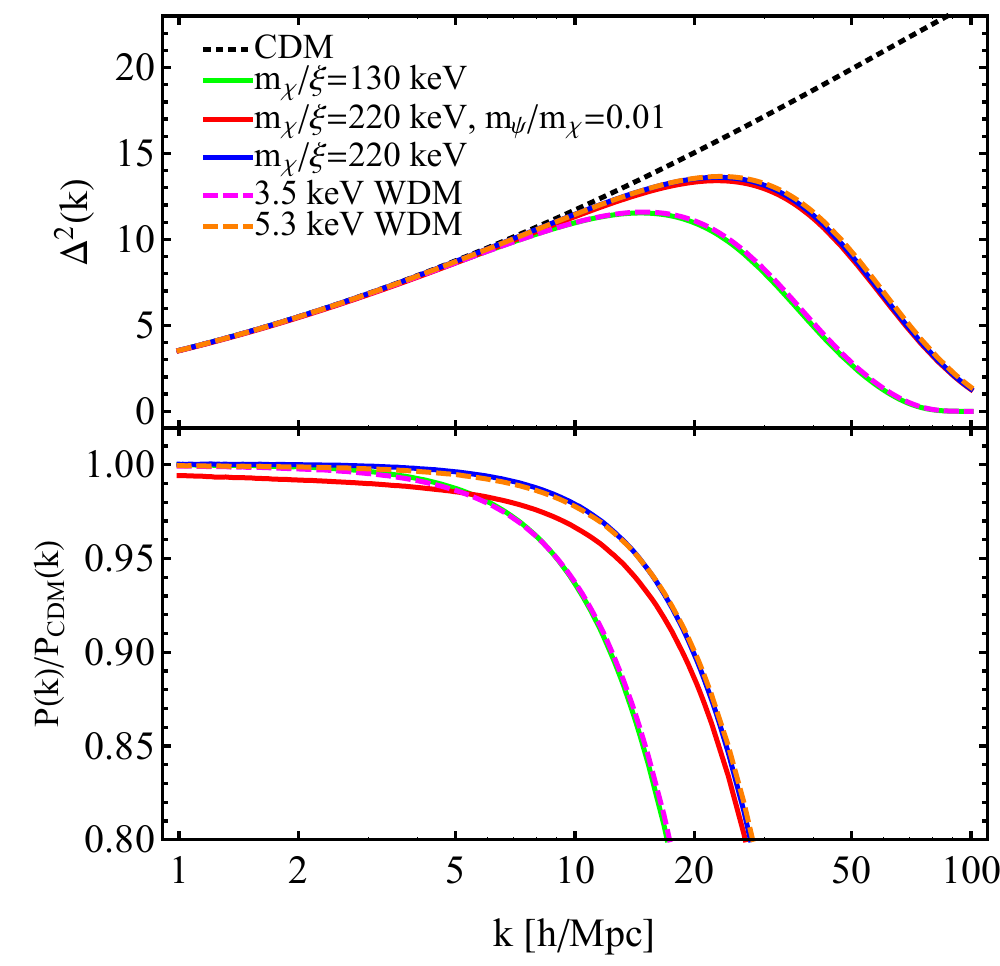}
\caption{Matter power spectra for different DM $\chi$ masses and different $m_\psi/m_\chi$ and $\langle\sigma v\rangle$ ($m_\psi=0$, with one exception). Also shown are the $3.5$~keV and $5.3$~keV WDM spectra (saturating the current conservative and aggressive Lyman-$\alpha$ bound~\cite{Irsic:2017ixq}) .}
\label{fig:MPS}
\end{figure}

The IAnDM model may also affect MPS in various ways. For the parameter range that leads to observable effects (light $\chi$), the dominating factor is the free-streaming of $\chi$ following its freeze-out. Although IAnDM freezes out as a cold relic, with sufficiently small mass (e.g. $\mathcal{O}(100)$ keV), it decouples late enough to sustain a non-negligible free-streaming effect for LSS (also see Ref.~\cite{Das:2010ts}). As shown in the later Fig.~\ref{fig:MPS}, at leading order, the MPS from IAnDM resembles that due to a thermal WDM model with a corresponding $m_{\rm WDM}$, which shows suppression at a much larger scale (kpc to Mpc) compared to a typical WIMP clump (pc)~\cite{Loeb:2005pm,Bertschinger:2006nq}, thus, it is subject to current Lyman-$\alpha$ constraints. The $\chi-\psi$ scattering may induce a competing suppression through dark acoustic oscillation~\cite{Loeb:2005pm,Bertschinger:2006nq}, but we have checked that with a massive $Z'$ and $\xi<1$, the $\chi-\psi$ kinetic decoupling occurs early enough and does not affect MPS effectively \footnote{For example with $\xi=0.1$, $m_\chi=1$~keV and $g_\chi=0.5$, the $m_{Z'}$ to reproduce thermal freeze-out $\langle\sigma v\rangle/\xi$ is $80$~MeV, and kinetic decoupling $T_\text{kd}=2.4$~keV from Eq.~2 of~\cite{Huo:2017vef} happens earlier than chemical freeze-out $T_f=1$~keV. This is due to a much colder dark sector, where the DR $\psi$ number density is $\xi^3$ suppressed. Therefore in our numerical calculation we have ignored the $\chi-\psi$ scattering.}. A correspondence between IAnDM and WDM can be derived analytically by matching the free-streaming velocity, $v_{\rm FS}(a)$, in the two models, which plays an important role in perturbation Boltzmann equations~\cite{Ma:1995ey} (this is similar to the correspondence between the sterile neutrino case and thermal WDM~\cite{Viel:2005qj}). In the nonrelativistic limit, $v_{\text{FS}}(a)=\frac{p}{E}\to\frac{\tilde{p}}{m}\frac{1}{a}$, where the tilde $(\tilde{})$ denotes comoving quantities. Therefore, if the $v_\text{FS}$'s match today ($a=1$), they also match at earlier times relevant to structure formation. Taking into account the specifics around freeze-out in IAnDM and WDM models, we find the following at $a=1$:  $\tilde{p}_\text{WDM}=(\frac{2\pi^2}{3\zeta(3)}\frac{\rho_{\text{WDM}}}{m_\text{WDM}})^{\frac{1}{3}}$, $\tilde{p}_\chi=\sqrt{3\hat{x}_f}\xi T_{\gamma}$, where $T_{\gamma}$ is $T_{\rm CMB}$ today. Equating $\frac{\tilde{p}}{m}$ in the two models, we find the correspondence
\beq
m_\text{WDM}=\Big(\frac{2\pi^2\rho_{\text{WDM}}}{3\zeta(3)}\Big)^{\frac{1}{4}}\Big(\frac{m_\chi/\xi}{T_{\gamma}\sqrt{3\hat{x}_f}}\Big)^{\frac{3}{4}}
=0.07\Big(\frac{m_\chi/\xi}{\text{keV}}\Big)^{\frac{3}{4}}~\text{keV},
\label{eq:FreeStreamingV}
\eeq
where we take $\rho_{\text{WDM}}=\rho_{\rm DM,0}$, $\hat{x}_f=10$.

\stepcounter{sec}\label{sec: numerics}
{\bf \Roman{sec}. Numerical Studies\;}

We further perform a numerical analysis by solving the perturbation Boltzmann equations for the evolution of the $\chi, \psi$ system using \texttt{camb}~\cite{Lewis:1999bs}. The general formulation follows Ref.~\cite{Ma:1995ey} with the choice of a synchronous gauge. The dominant DM, $\chi$, follows the standard treatment while allowing the DM free-streaming effect. As $\Omega_\chi$ will slowly decrease due to annihilation, for fixed $\langle\sigma v\rangle$, we numerically iterate to align it with the CDM value at matter-radiation equality. The initial perturbation of the annihilation product, $\psi$, inherits that of $\chi$ (CDM), while its later evolution follows the same perturbation multipole expansion as the massless or massive neutrinos. Therefore, for the partition function of both $\chi$ and $\psi$, $f=f(\tilde{p})(1+\Psi)=f(\tilde{p})(1+\Psi(\tau,x^i,p,n_i))$, i.e., the unperturbed $f(\tilde{p})$ is corrected by the small fractional perturbation $\Psi$. The perturbative Boltzmann equation for the multipole $\ell$ is
\begin{eqnarray}\label{massive neutrino}
\partial_\tau\Psi_0&=&-\frac{\tilde{p}}{\tilde{E}}k\Psi_1+\frac{1}{6}\partial_\tau h\frac{d\ln f(\tilde{p})}{d\ln\tilde{p}},\\
\partial_\tau\Psi_1&=&\frac{\tilde{p}}{\tilde{E}}k\left(\Psi_0-\frac{2}{3}\Psi_2\right),\\
\partial_\tau\Psi_2&=&\frac{\tilde{p}}{\tilde{E}}k\left(\frac{2}{5}\Psi_1-\frac{3}{5}\Psi_3\right)\\
&&-\left(\frac{1}{15}\partial_\tau h+\frac{2}{5}\partial_\tau\eta\right)\frac{d\ln f(\tilde{p})}{d\ln\tilde{p}},\\
\partial_\tau\Psi_\ell&=&\frac{\tilde{p}}{\tilde{E}}\frac{k}{2\ell+1}\bigg(\ell\Psi_{\ell-1}-(\ell+1)\Psi_{\ell+1}\bigg)\enspace\ell\geq3,\\
\partial_\tau\Psi_\ell&=&\frac{\tilde{p}}{\tilde{E}}k\Psi_{\ell-1}+\frac{\ell+1}{\tau}\Psi_\ell\qquad\text{as truncation}.
\end{eqnarray}
Here we choose the synchronous gauge $g_{\mu\nu}=a^2(^{-1}_{\quad\delta_{ij}+h_{ij}})$, and $h_{ij}(\vec{x})=\int\frac{d^3k}{(2\pi)^{3/2}}e^{i\vec{k}\cdot\vec{x}}\Big(\hat{k}_i\hat{k}_jh(\vec{k})+\big(\hat{k}_i\hat{k}_j-\frac{1}{3}\delta_{ij}\big)6\eta(\vec{k})\Big)$,
which gives the definition of the metric perturbation of $h$ and $\eta$.

However, there are key differences between neutrinos and the $\psi$ produced from DM annihilation. First, unlike neutrinos, the unperturbed partition function $f(\tilde{p})$ of $\psi$ is not thermal; rather, it is given by the annihilation process with a fixed physical momentum (determined by $m_\chi$, $m_\psi$). The $f(\tilde{p})$ feeds into the perturbation equations through the factor of $\frac{\partial\ln f(\tilde{p})}{\partial\ln \tilde{p}}$, which can be replaced by $-4$ for massless neutrinos. For our nonthermally produced $\psi$, this factor can be calculated in the same way as in Ref.~\cite{Huo:2011nz}, which at leading order yields $\frac{\partial\ln f(\tilde{p})}{\partial\ln \tilde{p}}=-\frac{9}{2}+\frac{3\bar{p}}{2\bar{\rho}}$, where $\bar{p}, \bar{\rho}$ are the background pressure and energy density. Numerically, we also include a small $\mathcal{O}(\frac{\Gamma_{\rm ann}}{H})$ correction to this factor. Moreover, the binning of $\psi$ in the comoving momentum space is different from that of the neutrino: for a fixed $a$, only the modes with $\tilde{p}<a\sqrt{m_\chi^2-m_\psi^2}$ contribute. We choose the standard cosmology parameters: $h=0.678$, $\Omega_\chi h^2\approx\Omega_{\rm DM}h^2=0.1186$, $\Omega_bh^2=0.02226$, etc.~\cite{Ade:2015xua}

{\bf{Numerical Results: The CMB Signatures}\;}

\begin{figure}[t!]
\includegraphics[scale=0.5]{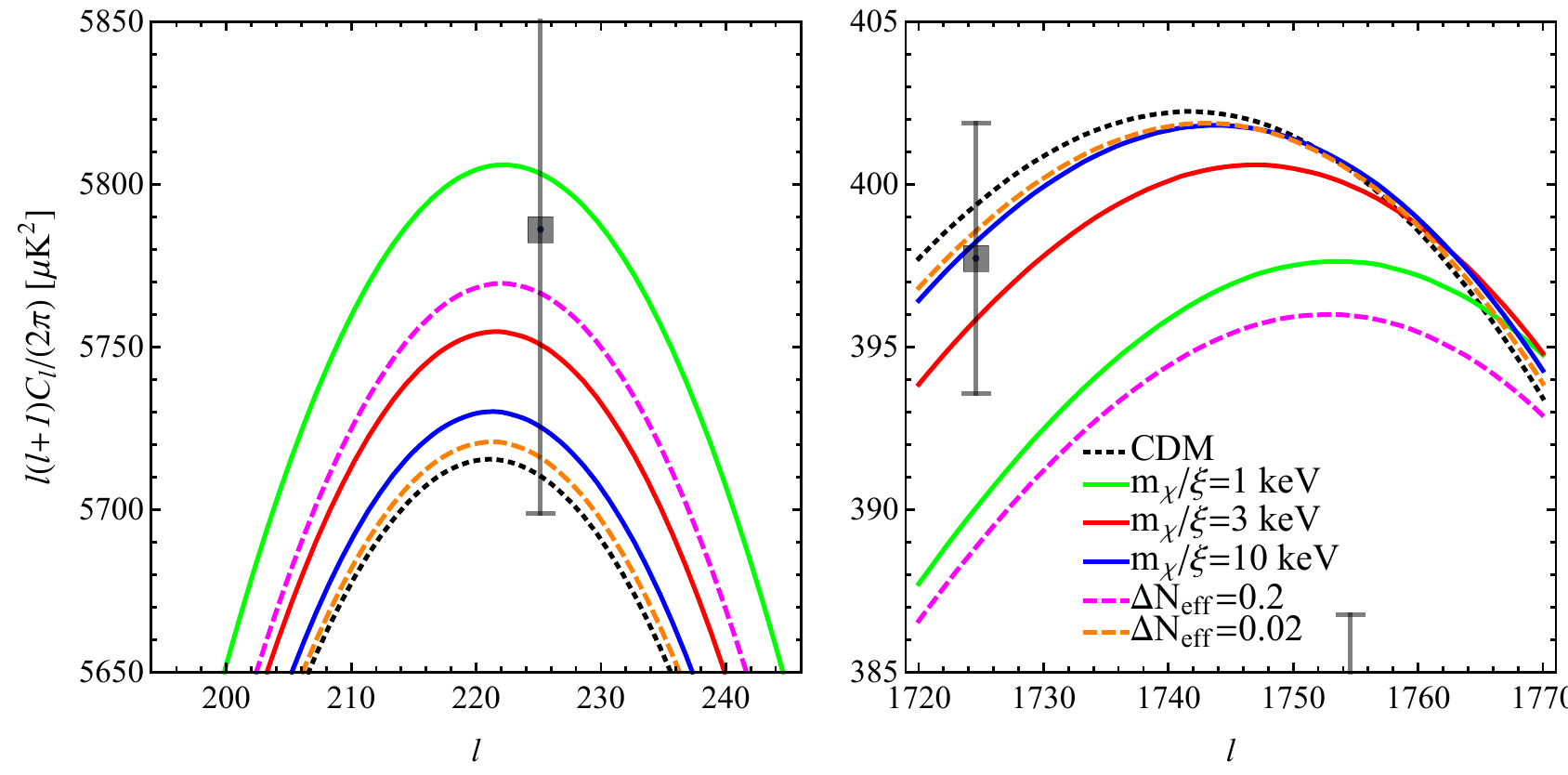}
\caption{An illustration of CMB signals of IAnDM with fixed cosmological parameters (see text). Also shown are the spectra with CDM and $\Delta N_\text{eff}=0.02,~0.2$ for comparison, as well as the binned error bars of Planck data. \emph{Left panel}: TT mode around the first acoustic peak. \emph{Right panel}: TT mode around the sixth acoustic peak. Note that this should not be used to extract phase shift since $\theta_*,\theta_D$ and $z_{\rm eq}$ are not fitted to observations.}
\label{fig:Low&HighL}
\end{figure}
In Fig.~\ref{fig:Low&HighL} we plot the zoomed-in CMB TT spectra around the first and the sixth acoustic peaks, for a set of $m_\chi/\xi$ choices, in comparison with CDM and standard $\Delta N_{\rm eff}$ models. As expected based on our analytic estimate (Eq.~\ref{eq:deltaNeff}), the effects from $\psi$ injection at leading order resemble those from standard $\Delta N_{\rm eff}$, which as shown below are confirmed by numerical studies.  However, as we will explain, there are key differences between a genuine $\Delta N_\text{eff}$ and the $\Delta N_{\rm eff}$-like effect caused by nonthermal $\psi$.

The most notable effect on the CMB spectrum due to $\Delta N_\text{eff}$ or nonthermal $\psi$ in IAnDM is on the heights of the acoustic peaks. In both models, the change to $H$ due to additional radiation energy density leads to an increase in the height of the first peak (i.e., early integrated Sachs-Wolfe (ISW) effect), and a decrease in that of the higher peaks (i.e., enhanced Silk damping). However, the heights of CMB acoustic peaks from IAnDM do not align with those due to a fixed standard $\Delta N_\text{eff}$. As expected (Eq.~\ref{eq:deltaNeff} and Fig.~\ref{fig:FreezeOutDeltaNeff}), $\Delta N_{\rm eff, nth}$ varies over time as $\rho_{\psi, \rm{nth}}$ accumulates while $\chi$ continues to annihilate. This can be seen by comparing the amplitudes of the CMB spectrum shown in Fig.~\ref{fig:Low&HighL}, where higher $\ell$ corresponds to an earlier time (when the mode reenters the horizon): consider a fixed $m_\chi/\xi=$1 keV - around the sixth peak the resulting CMB TT spectrum roughly aligns with that associated with $\Delta N_\text{eff}\approx0.12$; while around the first peak, it aligns with a larger $\Delta N_\text{eff}\approx0.35$.

The standard $\Delta N_{\rm eff}$ also reveals itself via a unique phase shift of the high-$\ell$ acoustic peaks due to neutrinolike modes that induce a significant anisotropic stress and propagate faster than the sound speed (free-streaming)~\cite{Bashinsky:2003tk, Baumann:2015rya}, which cannot be caused by other standard physics \footnote{Such an effect from SM neutrinos was recently detected with Planck data~\cite{Follin:2015hya}.}. IAnDM induces a phase shift as well, yet with a dramatic difference compared to that caused by standard $\Delta N_{\rm eff}$. We found that when adjusting $h,~\Omega_{\rm DM}h^2,~Y_{\rm He}$ to match the CMB observables of $\theta_*$, $\theta_D$, and $z_{\rm eq}$~\cite{Ade:2015xua}, in the IAnDM model the locations of the acoustic peaks shift towards high $\ell$'s, \textit{opposite} to the direction expected from standard free-streaming $\Delta N_\text{eff}$~\cite{Bashinsky:2003tk}. Such an effect resembles that due to fluid-like DR species~\cite{Chacko:2015noa, Baumann:2015rya, Brust:2017nmv}, yet for different reasons. Even though it is free-streaming, unlike neutrinos, the nonthermal $\psi$ dominantly produced at late times \textit{inherits a negligible initial anisotropic stress from $\chi$}, and consequently would not induce an additional phase shift along the same direction as neutrinos \footnote{Although $\psi$ may develop a larger anisotropic stress after production, our numerical analysis shows that its effect on CMB phase shift is subdominant to what we discussed in this paper.}. However, such $\psi$ does contribute to late-time radiation energy density, and thus reduces the energy fraction of the free-streaming SM neutrinos. As noted in Ref.~\cite{Chacko:2015noa} such a reduction (due to fluid-like DR or nonthermal $\psi$ here) leads to a phase shift along the direction opposite to that due to standard $\Delta N_\text{eff}$. Furthermore, just as with peak heights, there is scale dependence in phase shifts, which is a higher-order effect that may distinguish IAnDM from fluid-like DR (see Ref.~\cite{Choi:2018gho} for a related recent work) \footnote{Qualitatively the novel effects on CMB anisotropy spectrum we discussed here may also originate from the scenario of DM decaying to DR ~\cite{Huo:2011nz, Audren:2014bca, Poulin:2016nat}. However, for decaying DM with long enough lifetime to be cosmologically stable and satisfy existing constraints, the amount of DR produced during CMB epoch is tiny and the resultant signal strength is expected to be negligible.}.

{\bf{Numerical Results: The Matter Power Spectra}\;}

In Fig.~\ref{fig:MPS}, we present the MPS in the scenario of IAnDM, in comparison with other familiar new physics scenarios: CDM and WDM. Compared to the standard CDM prediction, all these new models cause a suppression of MPS relative to the standard cosmology for $k\gtrsim 1~h/\text{Mpc}$. The upper panel $\Delta^2(k)=\frac{k^3}{2\pi^2}P(k)$~\cite{Dodelson:2003ft} shows that the IAnDM spectra well overlap with WDM spectra with a corresponding mass estimated by Eq.~\ref{eq:FreeStreamingV}. Nevertheless, the zoomed-in bottom panel of Fig.~\ref{fig:MPS} shows slight differences between IAnDM and WDM spectra, allowing a massive $\psi$ (still requiring $m_{\psi}\ll m_\chi$ to be consistent with our starting assumption on cosmology). The suppression on small scales is more pronounced for a massive $\psi$, which can be understood as a massive $\psi$ contributes more energy density at the structure formation time than a massless $\psi$, boosting the free-streaming effect.

\stepcounter{sec}
{\bf \Roman{sec}. Summary and Discussion\;}
\begin{figure}[t!]
\includegraphics[scale=0.5]{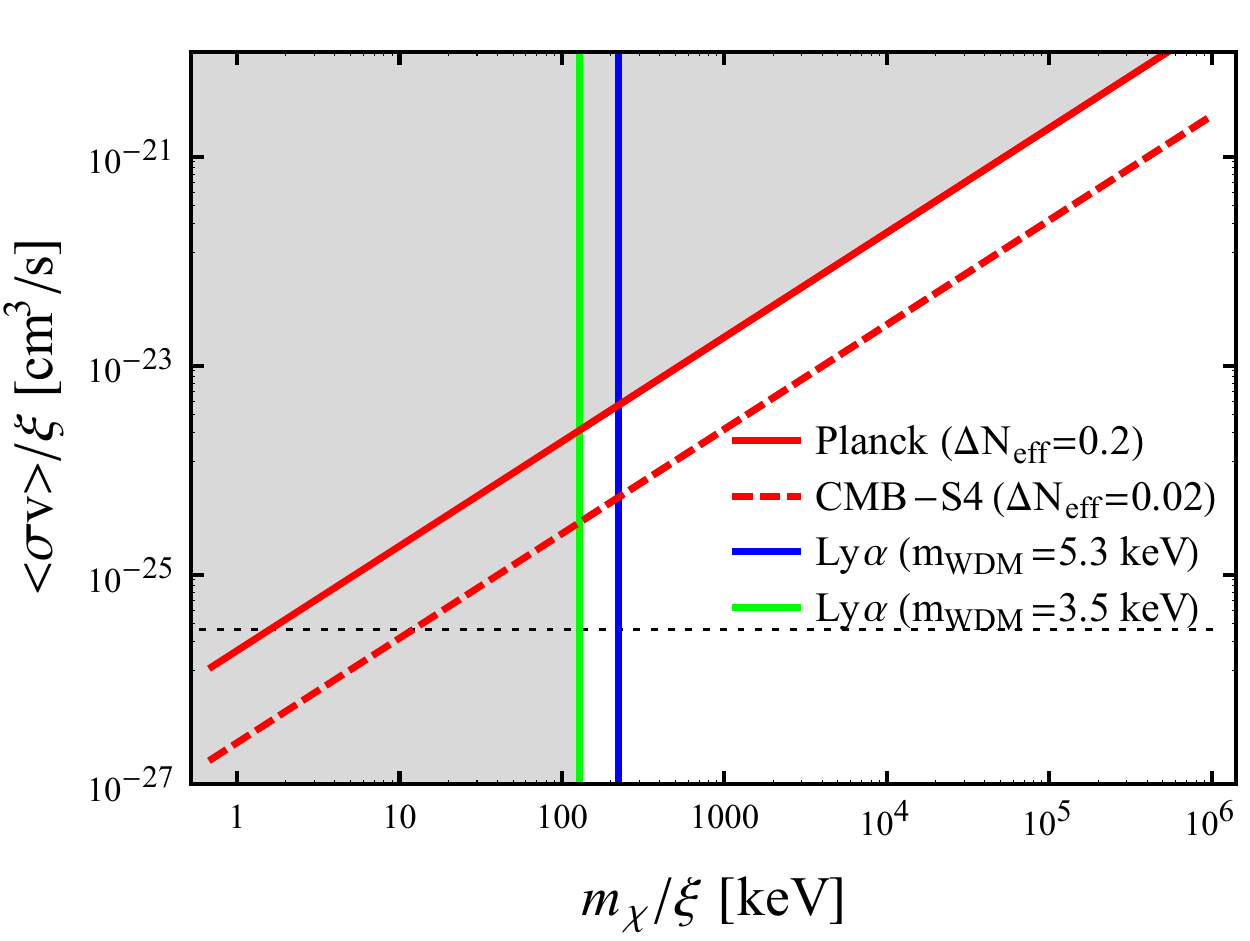}
\caption{The (projected) constraints on IAnDM from Planck and CMB-S4 (for details, see text), and from Lyman-$\alpha$ measurements ($m_\psi=0$, fixed $\langle\sigma v\rangle$); the shaded region is excluded. We assume all DM to be $\chi$, but allow mechanisms other than thermal freeze-out for its production. The axes are labeled by taking into account the rescaling factor $\xi=T_{\rm DM}/T_{\rm SM}$ at freeze-out time. Dashed line is a reference to standard thermal freeze-out cross section.}
\label{fig:SigmavVSMchi}
\end{figure}

With our understanding of the IAnDM effects on the CMB and the MPS, we can derive constraints based on current data and project sensitivities for future experiments.
The results are summarized in Fig.~\ref{fig:SigmavVSMchi}, where we label the axes with the scaling factor $\xi$ to account for the generic colder dark sector possibility. The CMB constraint is calculated by Fisher forecast, which agrees well with the estimate by comparing the (projected) experimental sensitivity limit on $\Delta N_\text{eff}$ and our analytical formula Eq.~\ref{eq:deltaNeff}. On the other hand, the MPS constraint is extracted by comparing the DM free-streaming effect based on our Eq.~\ref{eq:FreeStreamingV}. Note that the main effect of IAnDM on the CMB depends on the ratio $\langle\sigma v\rangle/m_\chi$ (Eq.~\ref{eq:deltaNeff}). The current Planck (projected CMB-S4) sensitivity for fixed $\Delta N_\text{eff}$, $\sigma_{N_\text{eff}}\simeq0.2$~\cite{Brust:2017nmv, Ade:2015xua} ($\sigma_{N_\text{eff}}\simeq0.02$~\cite{Abazajian:2016yjj, Wu:2014hta, Errard:2015cxa, Abazajian:2013oma}), gives a bound of $m_\chi/\xi\gtrsim1.6$~keV ($m_\chi/\xi\gtrsim12$~keV), assuming standard thermal $\langle\sigma v\rangle/\xi$. In contrast, the main effect of IAnDM on MPS only depends on $m_\chi/\xi$ (Eq.~\ref{eq:FreeStreamingV}), insensitive to $\langle\sigma v\rangle$. The current Lyman-$\alpha$ observation constrains IAnDM MPS: the $2\sigma$ aggressive bound ($m\gtrsim5.3$~keV for WDM) corresponds to a bound of $m_\chi/\xi\gtrsim220$~keV, and the conservative bound ($m\gtrsim3.5$~keV for WDM) corresponds to a bound of $m_\chi/\xi\gtrsim130$~keV.

Note that although in this work we considered a particular thermal IAnDM model as a simple example, the inferred novel phenomenology applies to broader possibilities of IAnDM models allowing an enhanced annihilation cross section relative to the standard thermal value. Such possibilities have been well considered in the context of the familiar visibly annihilating DM, including: Sommerfeld enhancement~\cite{ArkaniHamed:2008qn, Feng:2010zp, Binder:2017lkj} (automatic in IAnDM if the DR is a light mediator instead of a fermion), enhancement due to nonstandard cosmology, and nonthermal production of DM~\cite{Gelmini:2006pw, Fan:2014zua, Erickcek:2015jza}.

Astrophysical and cosmological observation is stepping into a high-precision era, which enables us to probe well-motivated DM models that are beyond the reach of conventional DM detections. In this article we demonstrate a representative example by investigating the novel phenomenology from the minimal and generic scenario of Invisibly Annihilating DM (IAnDM). The smoking-gun signature of this large class of models includes a correlated combination of scale-dependent, fluid-like $\Delta N_{\rm eff}$ in the CMB spectra and a characteristic matter power spectrum that resembles WDM. The current data constrain the IAnDM with masses up to $\sim 200$ keV, while future experiments can be sensitive to larger masses if the DM annihilation cross section is enhanced relative to the standard thermal value. These findings motivate new dedicated analyses to optimize the potential for discovering DM residing in a hidden sector.

{\it Acknowledgments}:\\
We thank Simeon Bird, Raphael Flauger, Takemichi Okui, Zhen Pan, Yuhsin Tsai and Hai-Bo Yu for discussion. The authors are supported in part by the US Department of Energy (DOE) grant DE-SC0008541.
\bibliography{refs_222}

\end{document}